\documentclass[aps,pra,twocolumn,showpacs,floatfix,groupedaddress]{revtex4}

\usepackage{graphicx}% Include figure files
\usepackage{dcolumn}% Align table columns on decimal point
\usepackage{bm}% bold math

\begin{document}

\title{Bose-Einstein Condensation Temperature of a Homogeneous Weakly Interacting
Bose Gas : PIMC study}

\author{Kwangsik Nho$^{*}$ and D. P. Landau}

\affiliation{Center for Simulational Physics, University of Georgia, Athens,
Georgia 30602}

\date{\today}

\begin{abstract}

Using a finite-temperature Path Integral Monte Carlo simulation
(PIMC) method and finite-size scaling, we have investigated the
interaction-induced shift of the phase transition temperature for
Bose-Einstein condensation of homogeneous weakly interacting Bose
gases in three dimensions, which is given by a proposed analytical
expression $T_{c} = T_{c}^{0}\{1 +
c_{1}an^{1/3}+[c'_{2}\ln(an^{1/3})+c''_{2}]a^{2}n^{2/3}
+O(a^{3}n)\}$, where $T_{c}^{0}$ is the critical temperature for
an ideal gas, $a$ is the $s$-wave scattering length, and $n$ is
the number density. We have used smaller number densities and more
time slices than in the previous PIMC simulations [Gruter {\it et
al.}, Phys. Rev. Lett. {\bf 79}, 3549 (1997)] in order to
understand the difference in the value of the coefficient $c_{1}$
between their results and the (apparently) other reliable results
in the literature. Our results show that
$\{(T_{c}-T_{c}^{0})/T_{c}^{0}\}/(an^{1/3})$ depends strongly on
the interaction strength $an^{1/3}$ while the previous PIMC results
are considerably flatter and smaller than our results. We obtain
$c_{1}$ = 1.32 $\pm$ 0.14, in agreement with results from recent
Monte Carlo methods of three-dimensional $O(2)$ scalar $\phi^{4}$
field theory and variational perturbation theory.

\end{abstract}

\pacs{05.30.Jp,03.75.Hh,02.70.Ss}

\maketitle

\section{Introduction}
Since the recent achievement of experimental observation of
Bose-Einstein Condensation (BEC) in magnetically trapped atomic
vapors\cite{ADB}, these inhomogeneous systems have attracted
considerable interest from experimental and theoretical sides. In
dilute or weakly interacting gases, the Gross-Pitaevskii (GP)
theory, a mean-field theory, for the condensate wave function has
been enormously successful in describing the extraordinary
properties of the condensate. (For a recent review of numerous
successful applications of mean field theories in Bose-Einstein
condensation in atomic gases see Ref. (2).) However, the
determination of the effect of repulsive interactions on the
transition temperature of a homogeneous dilute Bose gas at a fixed
density has had a long and controversial history\cite{LY, Stoof1,
Stoof2, Gruter, Huang, Baym1, Baym2, Baym3, Holzmann1, Holzmann2,
Holzmann3, Holzmann4, Arnold1, Arnold2, Arnold3, Ramos1, Ramos2,
Ramos3, Reppy, Wilkens, Alber, Kashurnikov, Braaten, Kleinert,
Sun, Kastening1, Kastening2}. This non-trivial problem has been
treated by different methods with different results. One of the
reasons for the multitude of results and methods stems from the
fact that the phase transition is second order, and perturbation
theory typically breaks down for physical quantities sensitive to
the collective long-wavelength modes close to the transition due
to infrared (IR) divergences. Moreover, the interaction, no matter
how small, changes the universality class of the phase transition
from the Gaussian complex-field model to that of the $XY$-model.

In the weak interaction (or dilute) limit, the strength of the
interatomic interactions can be characterized by the $s$-wave
scattering length $a$.  The shift $\Delta T_{c} \equiv
(T_{c}-T_{c}^{0})$ of the BEC transition temperature of a
homogeneous Bose gas away from its ideal-gas value
\begin{equation}
T_{c}^{0} = \frac{\hbar^{2}2\pi}{mk_{B}}[n/\zeta(\frac{3}{2})]^{2/3}
\end{equation}
behaves parametrically as
\begin{equation}
\frac{\Delta T_{c}}{T_{c}^{0}} \rightarrow c_{1}an^{1/3},
\label{old}
\end{equation}
where $m$ is the particle mass, $k_{B}$ is the Boltzmann constant,
$n$ is the number density, $\zeta(s)$ is the Riemann zeta
function, $\zeta(\frac{3}{2}) \approx 2.612$, and $c_{1}$ is a
numerical coefficient. Additionally, M. Holzmann {\it et al.}
\cite{Holzmann3} have argued that a logarithmic term appears at
order-$a^{2}$ in Eq. (\ref{old}), and they have shown that this
term is of the form $c'_{2}a^{2}n^{2/3}\ln(an^{1/3})$. They also
estimated the value of the numerical coefficient $c'_{2}$ using
large-$N_s$ arguments in the three-dimensional $O(N_s)$ field theory. 
Later, Arnold {\it et al.}\cite{Arnold3} were
able to show that the transition temperature for a dilute,
homogeneous, 3$D$ Bose gas can be expressed to leading order as
\begin{equation}
\frac{\Delta T_{c}}{T_{c}^{0}} = c_{1}an^{1/3}+[c'_{2}\ln(an^{1/3})
+c''_{2}]a^{2}n^{2/3}+O(a^{3}n),
\label{arnold}
\end{equation}
where $c'_{2} \simeq 19.7518$ can be calculated perturbatively,
whereas $c_{1}$ and $c''_{2}$ require non-perturbative techniques.
With the help of Monte Carlo data they estimated $c''_{2} \approx
75.7$; however, a strong debate concerns the values of the other
numerical coefficients, especially $c_{1}$ which has been computed
using various non-perturbative methods and Monte Carlo techniques.
There are numerous estimates for the parameter $c_{1}$ describing
the leading deviation of the BEC temperature due to a small
repulsive interaction, and the range of variation of different
predictions is from 0.34 (Ref. 6) to 4.66 (Ref. 21); these have been
summarized for instance in Refs. (10, 16, 30). It appears that the
most reliable results so far are obtained by Monte Carlo (MC)
simulations of three-dimensional $O(2)$ scalar $\phi^{4}$
field theory, $c_{1}$ = 1.29 $\pm$ 0.05 determined by Kashurnikov
{\it et al.}\cite{Kashurnikov} and $c_{1}$ = 1.32 $\pm$ 0.02
determined by Arnold {\it et al.}\cite{Arnold2}, and by
variational perturbation theory (VPT), $c_{1}$ = 1.27 $\pm$ 0.11
determined by Kastening\cite{Kastening2}. In particular,
Gruter {\it et al.}\cite{Gruter} investigated the
dependence of $\Delta T_{c}$ numerically using Path Integral Monte
Carlo methods. They obtained the value of $c_{1}$ = 0.34 $\pm$
0.06 after numerical extrapolation of the calculation to the limit
$a \rightarrow 0$. This value is about 4 times smaller than the
above results. Holzmann {\it et al.}\cite{Holzmann3} argued that
the difference in these results is attributable to a nonanalytic
structure, $a^{2}\ln a$, of the transition temperature in $a$, and
that this correction gives rise to a strong dependence on $a$,
even in the very dilute limit. Arnold {\it et al.} \cite{Arnold2}
believed that one likely problem with the PIMC simulation of
Gruter {\it et al.}\cite{Gruter} is inadequate system
size. Kashurnikov {\it et al.}\cite{Kashurnikov} stated that the
PIMC simulation did not reach the universality region because the
minimal value of $na^{3}$ was $~10^{-5}$.

In this paper, we attempt to understand the difference in the
value of the coefficient $c_{1}$ between results from PIMC and the
above results. We have used a finite-temperature path-integral
Monte Carlo (PIMC) method\cite{Ceperley} to study the dependence
on interaction strength of the phase transition temperature for
Bose-Einstein Condensation of weakly interacting Bose gases in
three dimension. Although our PIMC method is the same as that used
by Gruter {\it et al.}, we have used smaller number
densities and more time slices than in the previous PIMC
simulations, as motivated above.

\section{Simulation Method and Definition of the physical quantities}

We wish to study the problem of a quantum $N$-particle system in order to
compute the phase transition temperature $T_{c}$ for dilute or weakly
interacting Bose gases in three dimensions. We assume that the interparticle
interaction can be described by a positive scattering length $a$, equivalent
to the interaction of hard spheres of diameter $a$ (see, e.g., Refs. [2,3]).
The Hamiltonian for this system may be written as

\begin{equation}
H = -\frac{\hbar^{2}}{2m}\sum_{i=1}^{N} \nabla_{i}^{2} + \sum_{i<j}v(\mid
{\bf r}_{i}-{\bf r}_{j}\mid),
\end{equation}
where $v(r)$ is the hard-sphere potential defined by

\begin{eqnarray*}
v(r) & = & +\infty \hspace{1in}(r < a) \nonumber\\
     & = &  0      \hspace{1.2in}(r > a).
\end{eqnarray*}

The statistical mechanics of quantum systems is governed by the density matrix.
For a system of $N$ bosons at an inverse
temperature $\beta$, the Bose-symmetrized density matrix is given by

\begin{equation}
\rho_{B}({\bf R},{\bf R}';\beta) = \frac{1}{N!}\sum_{P}\rho({\bf R},P{\bf R}'
;\beta),
\label{eq-dm}
\end{equation}
where ${\bf R}$ and ${\bf R}'$ are two configurations of $N$ hard spheres.
$P$ denotes a permutation of particle labels among hard spheres, and $P{\bf R}$
is one such permutation. Evaluating the density matrix for interacting systems
at very low temperatures is complicated by the fact that the kinetic and
potential terms in the exponent of the density matrix cannot be separated.
We can aviod this problem by inserting $M-1$ intermediate configurations into
Eq.(\ref{eq-dm}) to obtain the path-integral formulation of the density matrix:

\begin{eqnarray}
\rho({\bf R},P{\bf R}';\beta) & = & \int\cdot\cdot\cdot\int d{\bf R}_{1}d{\bf
R}_{2} \cdot\cdot\cdot d{\bf R}_{M-1} \nonumber \\
                              & \times & \rho({\bf R},{\bf R}_{1};\tau)
\cdot\cdot\cdot\rho({\bf R}_{M-1},P{\bf R}';\tau),
\label{eq-dm2}
\end{eqnarray}
where $\beta$ = $1/k_{B}T$ and $\tau$ = $\beta/M$ is the imaginary
time step. The problem of evaluating the density matrix at a low
temperature $\beta^{-1}$ has been replaced by the problem of
multiple integration of density matrices at a higher temperature
$\tau^{-1}$. The sum over permutations in Eq.(\ref{eq-dm})
combined with the integration in Eq.(\ref{eq-dm2}) can be
evaluated in path-integral Monte Carlo (PIMC) by a stochastic
sampling of the discrete paths $\{{\bf R}, {\bf R}_{1}, {\bf
R}_{2}, \cdot\cdot\cdot, {\bf R}_{M-1}, P{\bf R}'\}$ using
multilevel Monte Carlo sampling\cite{Ceperley}, an extension of
the standard Metropolis method\cite{Metro}. The PIMC method is
essentially exact, the only necessary input being the
interparticle potential or equivalently the scattering length $a$.
In order to use Monte Carlo sampling, we
must first provide a pair-product form of the exact two-body
density matrices for the high-temperature density matrices that
appear in the integrand Eq.(\ref{eq-dm2}). We used the
high-temperature approximation for the hard-sphere propagator
derived by Cao and Berne\cite{Cao}. In order to choose the value
of $M$, we performed consistency checks by varying $M$ to see that
the results have converged. We used up to fifteen time-slices ($M =
15$). We performed PIMC in the continuum. The particles were confined 
to a cubic box with volume $V$ and edge length $L = V^{1/3}$, to which
periodic boundary conditions were applied. We employed the canonical
ensemble, {\it i.e.} in each simulation we fixed the temperature
$T$, the number of particles $N$, and the dimensions $L$ of the
simulation cell. 20,000 - 30,000 MC steps are required for equilibration 
depending on the number density. Statistical averages are collected from
30,000 MC steps after this. In each MC step, we attempted  50 - 800 trial moves
at each time slice. We performed these calculations partly on IBM SP2 and partly
on PC at the University of Georgia. The smallest-density simulations took about
350 CPU hours on IBM SP2.

The scaling functions for the condensate density and the superfluid density
are similar and imply that in the (dilute) interacting 3D Bose gas condensation
and superfluidity occur at precisely the same temperature. Superfluid density
can be calculated using the winding number ${\bf W}$ for simulations that have
periodic boundary conditions\cite{Pollock1}. Nonzero winding numbers occur 
when particles,
through a series of permutations, are permuted with periodic images of
themselves. The winding number is directly related to $\rho_{s}$, the
superfluid density. The superfluid density is given by\cite{Pollock1}

\begin{equation}
\frac{\rho_{s}}{\rho} = \frac{m}{\hbar^{2}}\frac{ \langle {\bf W}^{2} \rangle}
{3\beta N},
\end{equation}
where the winding number ${\bf W}$ is defined by

\begin{equation}
{\bf W} = \sum_{i=1}^{N}\int_{0}^{\beta}dt[\frac{d{\bf r}_{i}(t)}{dt}].
\end{equation}

In order to obtain an estimate of the superfluid transition temperature
$T_{c}$, we perform a finite-size scaling analysis.
Near the critical temperature $T_{c}$, the finite-size behavior
of the superfluid density obeys
the scaled form\cite{Pollock2}
\begin{equation}
\rho_{s}(t,L)/\rho = L^{-\pi/\nu}f(tL^{1/\nu}),
\label{eq-sc}
\end{equation}
where $\pi$ is the critical exponent of the bulk superfluid
fraction, $\rho_{s}(t)/\rho \sim t^{\pi}$, $t=(T-T_{c})/T_{c}$,
$\nu$ is the correlation length exponent, $\xi(t) \sim t^{-\nu}$,
and the universal function $f(tL^{1/\nu})$ must be analytic for
finite argument. It is assumed that $\pi = \nu$ since this is
indicated by experiment\cite{Grey}, renormalization-group
calculation\cite{Cle}, and the Josephson hyperscaling relation
$\pi = (d-2)\nu$. In obtaining an estimate of the critical
temperature $T_{c}$, we didnot use Eq. (\ref {eq-sc}) directly
because of large statistical errors in our data (see Fig 2(a)).
Instead we fit our data for different temperatures $T$ and several
values $N$ of the total number of particles to a linear form for
$f$
\begin{eqnarray}
N^{1/3}\rho_{s}(t,N)/\rho & = & f(tN^{1/3\nu})  \nonumber \\
                          & = & f(0)+f_{0}N^{1/3\nu}(T-T_{c})/T_{c},
\label{fite}
\end{eqnarray}
where the dimensionless $N^{1/3}$ has replaced the length $L$.
From the fitting, we determined four fitting parameters $f(0)$, 
$f_{0}$, $\nu$, and $T_{c}$.

\section{Simulation Results and Discussion}

The main goal of this paper is to determine the critical
temperature $T_{c}$ of a 3$D$ homogeneous system of hard-sphere
bosons by path-integral Monte Carlo simulations and finite-size
scaling. We have calculated the superfluid fraction
$\rho_{s}(T,N)/\rho$ for various number densities $n$ at a fixed
hard-sphere diameter $a$.

\begin{figure}[ht]
\begin{picture}(0,400)(0,0)
\put(-140,-20){\includegraphics{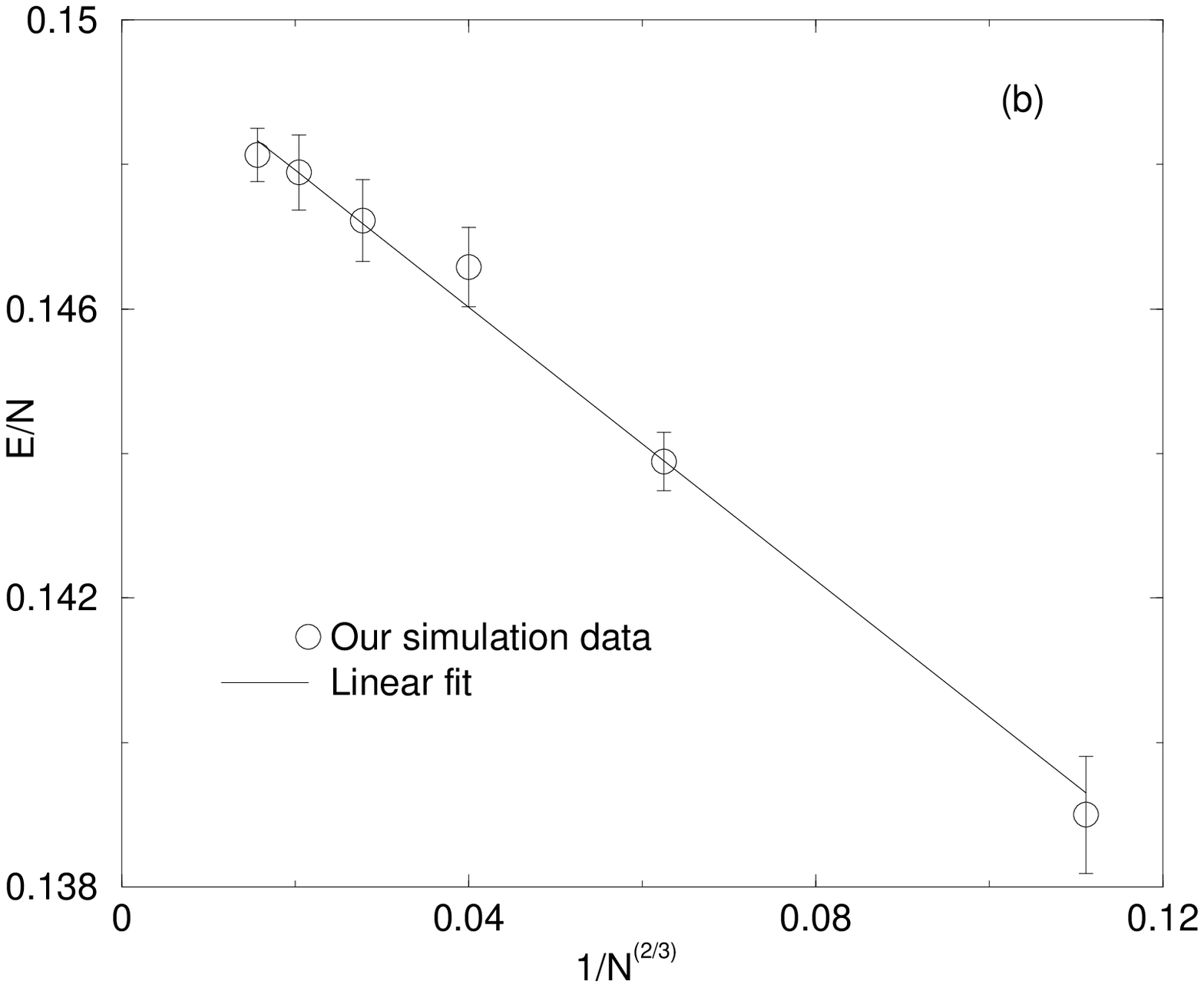}}
\put(-140,180){\includegraphics{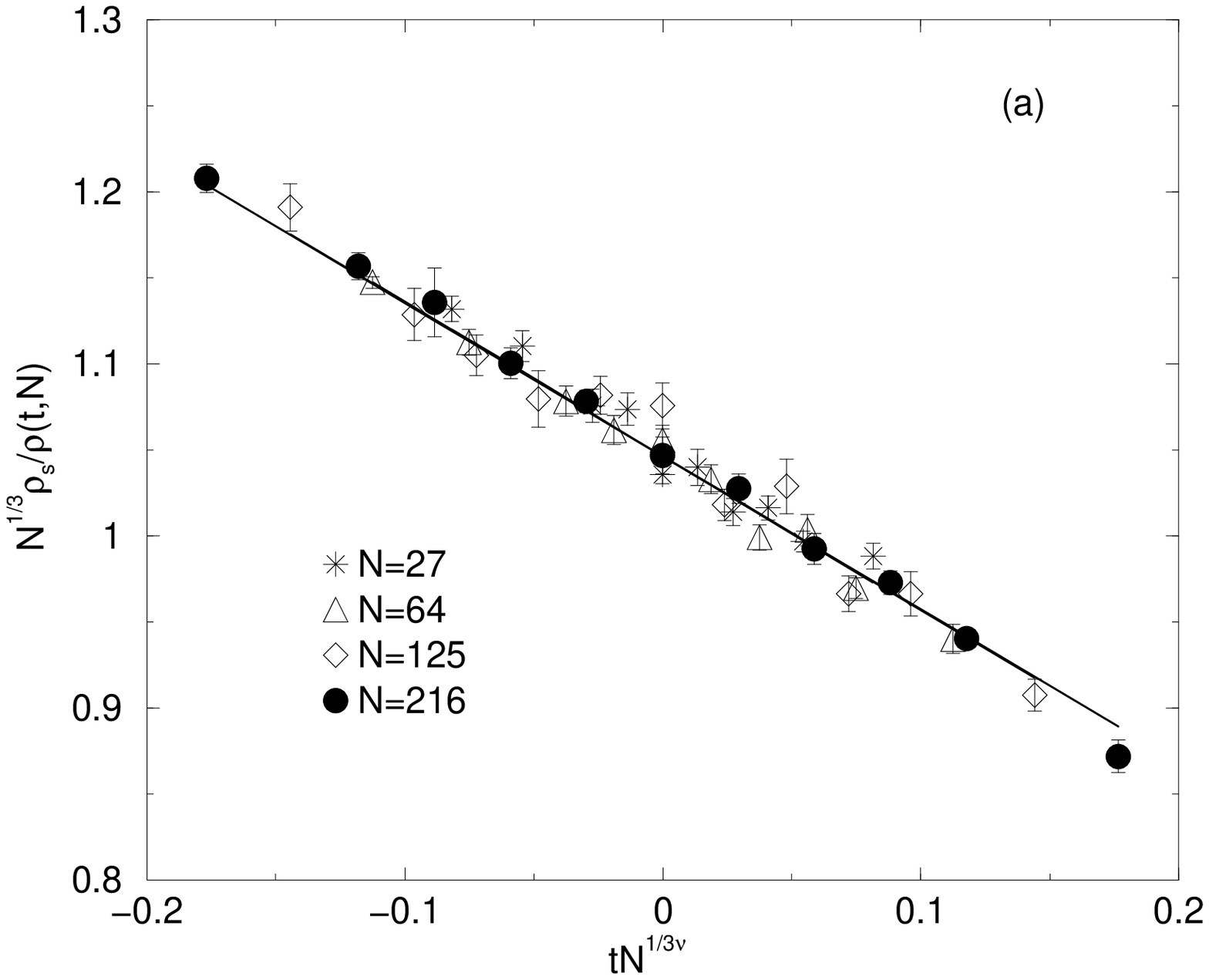}}
\end{picture}
\caption{Results for the noninteracting case. (a) The scaled
superfluid fraction $N^{1/3}\rho_{s}(t,N)/\rho$ as a function of
scaled temperatures $tN^{1/3\nu}$. The solid line is a linear fit
to Eq. (\ref{fite}). The best fit parameters are the critical
temperature of $T_{c}/T_{c}^{0}$= 1.000(1) and the correlation
length exponent of $\nu$ = 0.90(6). (b) The total energy density
as a function of the number of particles with a straight line fit
to Eq. (\ref{eq-e}). From the fit we obtain the energy density in
the thermodynamic limit ($N\rightarrow \infty$). } \label{fi-1}
\end{figure}

In order to check the program carefully, we determined the
critical temperature $T_{c}^{0}$ and the total energy density in
the noninteracting case, both of which are known exactly. Fig.
1(a) shows the scaled superfluid fraction
$N^{1/3}\rho_{s}(t,N)/\rho$, scaled temperarure $tN^{1/3\nu}$, and
the resulting linear fit (solid line) to Eq. (\ref{fite}) for the
noninteracting hard spheres. The best fit parameters are the
critical temperature of $T_{c}/T_{c}^{0}$= 1.000(1) and the
correlation length exponent of $\nu$ = 0.90(6), in agreement
with the theoretical exact values, namely $T_{c}/T_{c}^{0}$ = 1
and $\nu$ = 1.

From the expression for the specific heat $c(t,L)$,
\begin{equation}
c(t,L) = \frac{\partial E(t,L)}{\partial T},
\end{equation}
we obtain the energy density $E(t,L)$, via integration, up to a constant

\begin{equation}
E(t,L) = c(0,\infty)T + L^{(\alpha - 1)/\nu}T_{c}D(tL^{1/\nu}),
\end{equation}
where $t = (T-T_{c})/T_{c}$. Defining $dD(x)/dx = g(x)$, we write
the specific heat as follows :

\begin{equation}
c(t,L) = c(0,\infty) + L^{\alpha/\nu}g(tL^{1/\nu}),
\end{equation}
where $g(x)$ is a universal scaling function for the specific heat.
For $t \rightarrow 0$ we obtain
\begin{equation}
E(0,L) = E_{c} + E_{1}L^{(\alpha - 1)/\nu}.
\label{eq-e}
\end{equation}

\begin{figure}[ht]
\begin{picture}(0,400)(0,0)
\put(-140,-15){\includegraphics{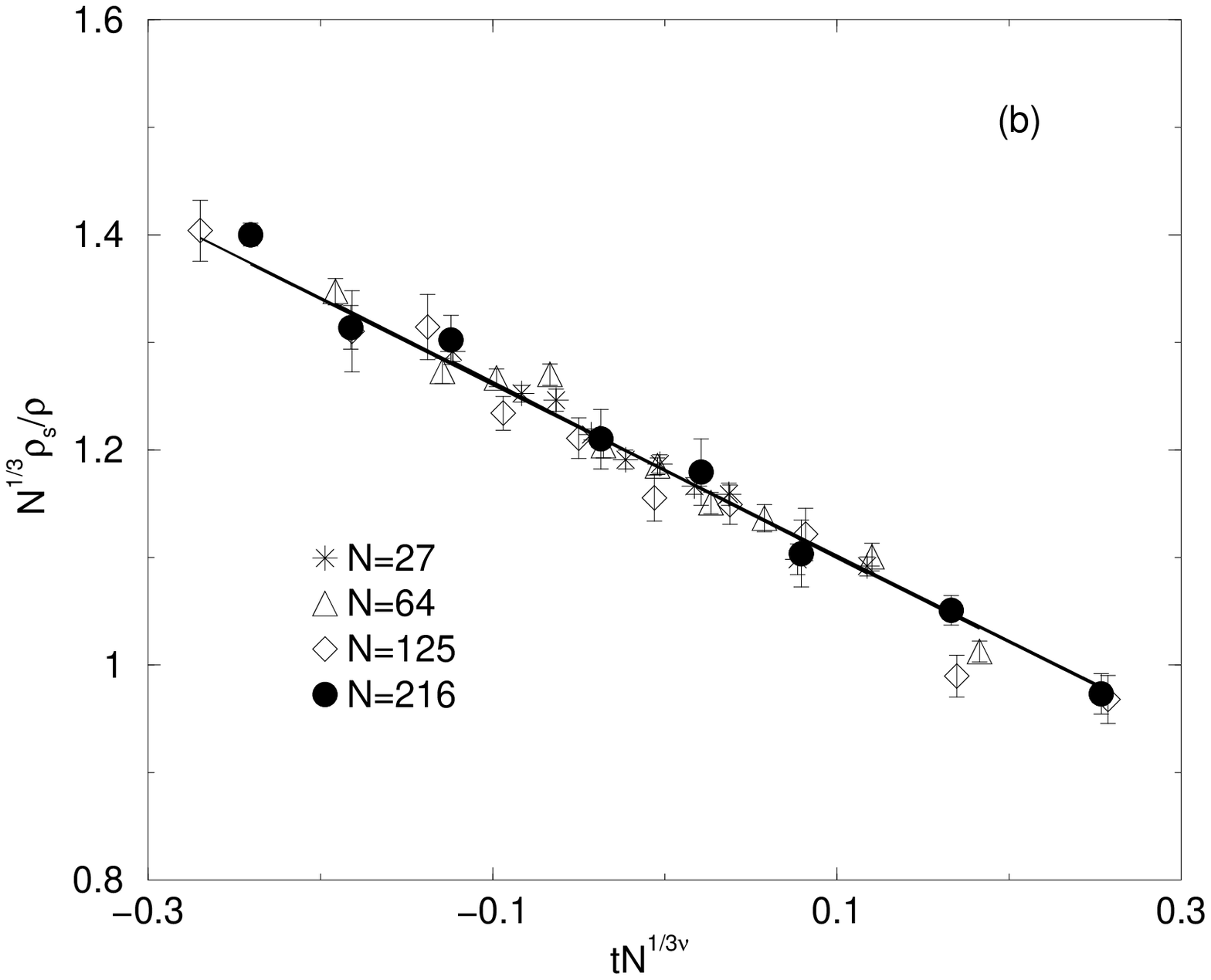}}
\put(-140,180){\includegraphics{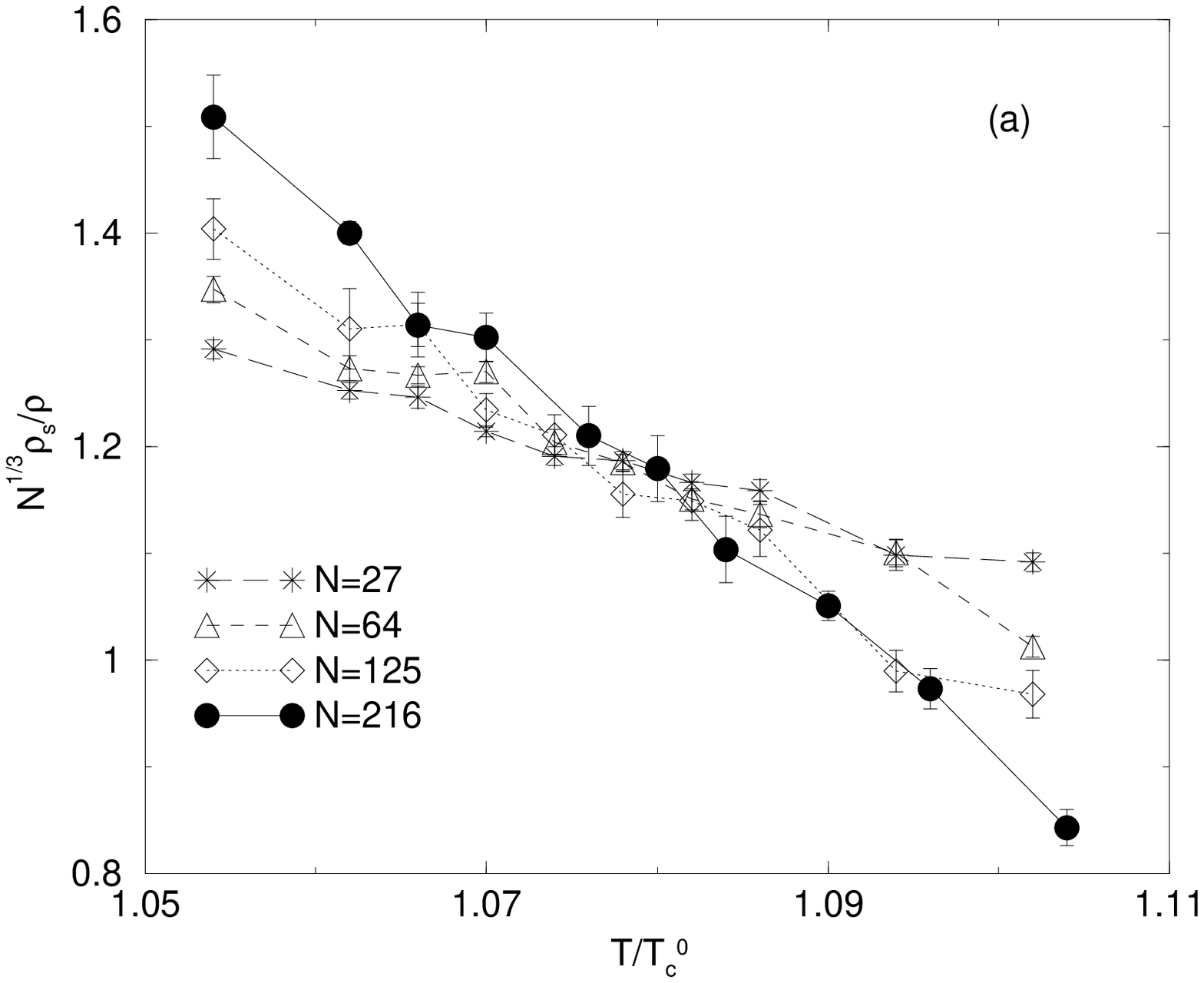}}
\end{picture}
\caption{Determination of the critical temperature $T_{c}$ at a
fixed number density $n$ = $5\times 10^{-3}$ for the interacting
case. (a) The scaled superfluid fraction
$N^{1/3}\rho_{s}(t,N)/\rho$ as a function of scaled temperatures
$T/T_{c}^{0}$. The four lines cross at the critical temperature.
(b) Result of our linear fit (solid straight line) to the data
from Eq. (\ref{fite}). Our estimated critical temperature is
$T_{c}/T^{0}_{c}$ = 1.078(1). } \label{fi-2}
\end{figure}

The result of the fit of the energy density data to Eq.
(\ref{eq-e}) is shown in Fig. 1(b). In the fit we used $\alpha =
-1 $ and $\nu = 1 $ for an ideal Bose gas. If we exclude the data
corresponding to the smallest lattice, we obtain an estimate for
the energy density in the thermodynamic limit ($N\rightarrow
\infty$) which agrees with the exactly known value within an
errorbar.

\begin{figure}
\begin{picture}(0,200)(0,0)
\put(-140,-20){\includegraphics{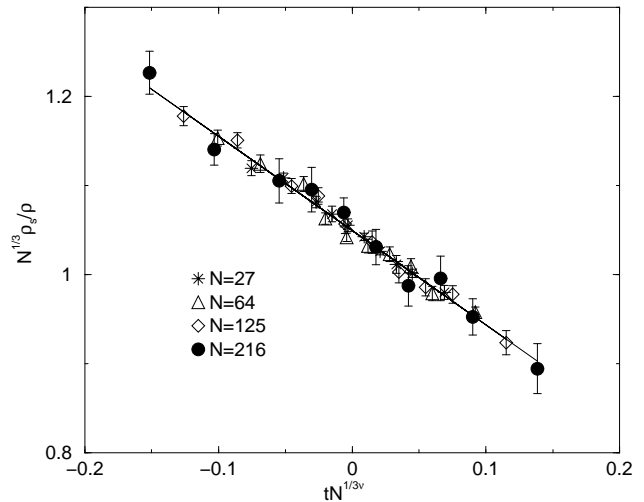}}
\end{picture}
\caption{ Scaled superfluid fraction $N^{1/3}\rho_{s}(t,N)/\rho$ vs 
$tN^{1/3\nu}$ at the smallest number density
$n$ = $1 \times 10^{-7}$ we used in this study. Our estimated critical 
temperature is $T_{c}/T^{0}_{c}$ = 1.0061(14).}
\label{fi-3}
\end{figure}

In the interacting case, we have calculated the superfluid
fraction and estimated the critical temperatures for various
hard-sphere diameters $a$ at a fixed number density. We can see
that the critical temperature approaches $T_{c}=T_{c}^{0}$ as the
hard-sphere diameter decreases (not shown), as expected.

In order to study the effect of interactions on the transition temparature,
we calculated the superfluid fraction $\rho_{s}(t,N)/\rho$ and determined
the critical temperatures $T_{c}(n)$ for various number densities
$1 \times 10^{-7} \leq n \leq 5 \times 10^{-3}$.

\begin{figure}[ht]
\begin{picture}(0,360)(0,0)
\put(-140,-32){\includegraphics{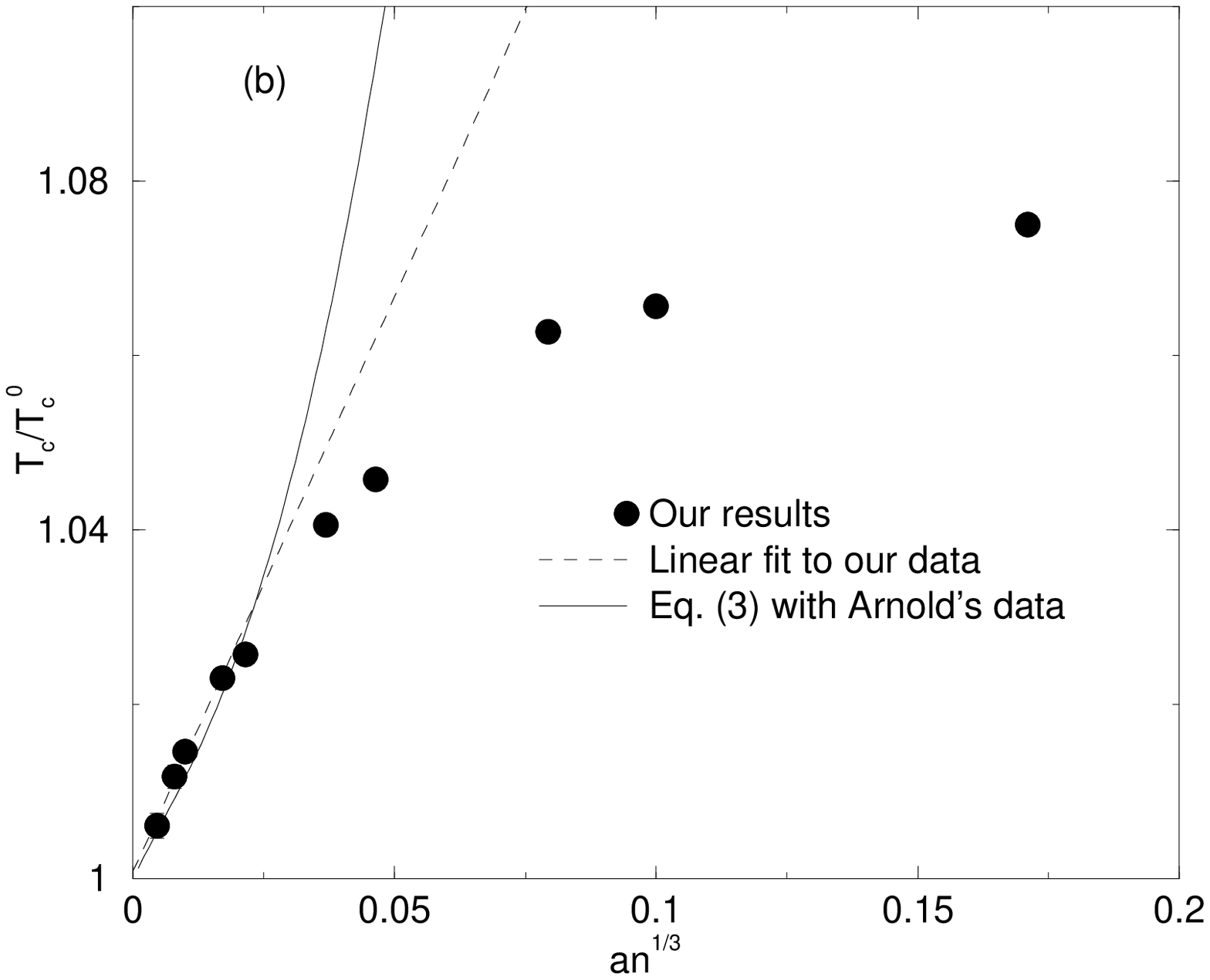}}
\put(-140,140){\includegraphics{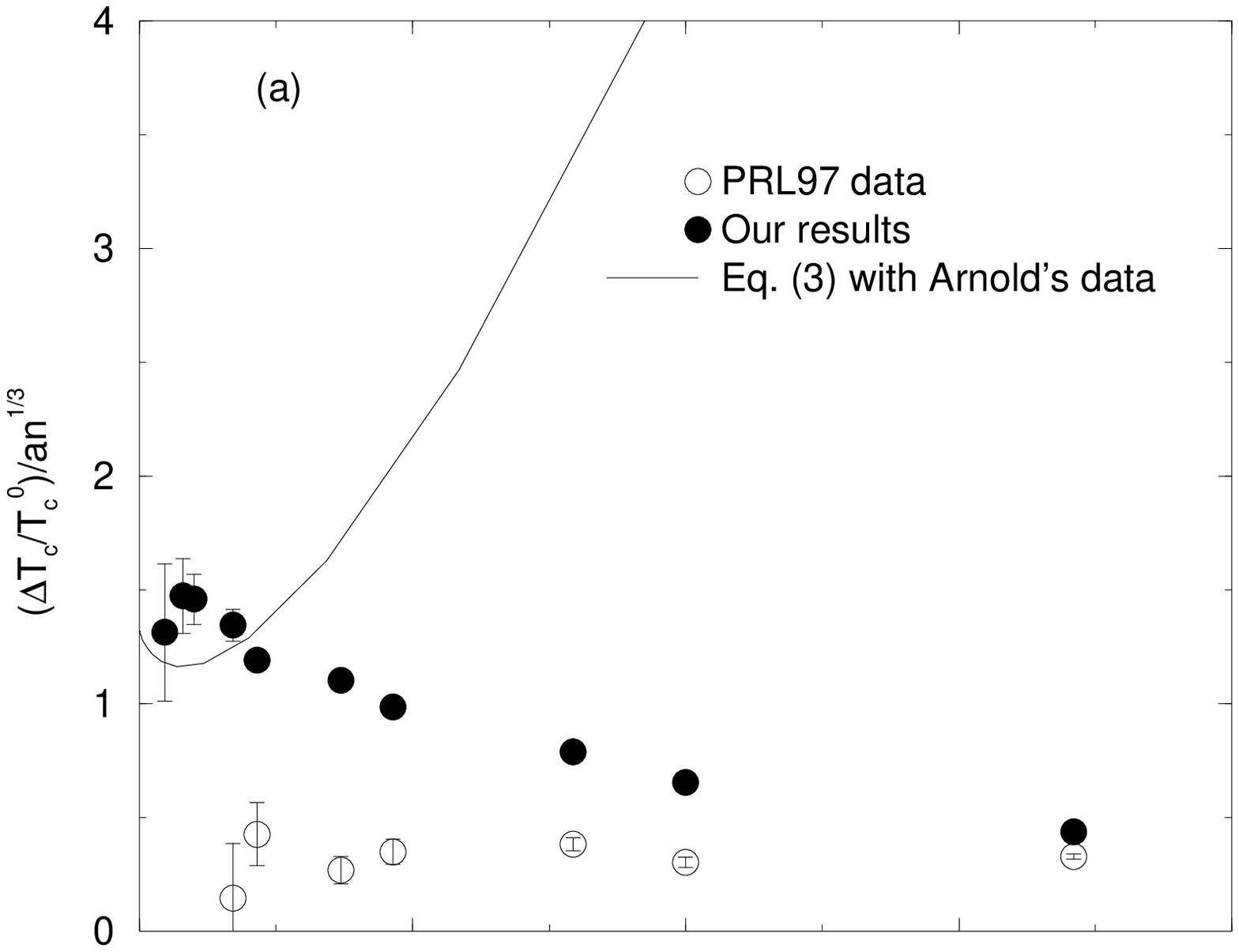}}
\end{picture}

\caption{ Our calculated dependence of the interaction-induced shift of
the transition temperature of a dilute homogeneous Bose gas
on the scattering length $an^{1/3}$: (a) $(\Delta T_{c}/T^{0}_{c})/an^{1/3}$
(b) $T_{c}/T_{c}^{0}$. Whenever not shown, the error bars in our
simulation data are smaller than the symbol sizes.
We also plotted the PRL97 data (Open circles) and
Eq. (\ref{arnold}) using Arnold's results\cite{Arnold2} (Solid line).
(a) $(\Delta T_{c}/T^{0}_{c})/an^{1/3}$ depends on
the interaction strength $an^{1/3}$ while the PRL97 data are considerably 
flatter and smaller than our results.
(b) The linear behavior is seen only at small value of $an^{1/3}$.
The dashed straight line is a fit to our four smallest data points.
We found that the slope ($c_{1}$) is 1.32(14).}
\label{fi-4}
\end{figure}

Gruter, Ceperley, and Laloe\cite{Gruter}
investigated the dependence of $\Delta T_{c}$ numerically using
PIMC (to be referred to here as PRL97). In order to compare our
estimated critical temperature with PRL97 data, we have calculated
the superfluid fraction $\rho_{s}/\rho (T,N)$ for $N$ = 27, 64,
125, and 216 as a function of temperature $T/T_{c}^{0}$ at a fixed
number density $n$ = $5\times 10^{-3}$. Fig. 2 shows our data
(Fig. 2(a)) and our result of a linear fit (solid straight line)
to the data (Fig. 2(b)) using Eq. (\ref{fite}). Our calculated 
superfluid fractions $\rho_{s}(t,N)/\rho$ are significantly
larger than PRL97 data (cf. Fig. 1 of Ref. 6) and our estimated
critical temperature is $T_{c}/T^{0}_{c}$ = 1.078(1), which is
higher than PRL97 data, $T_{c}/T^{0}_{c}$ = 1.057(2) 
(cf. Fig. 2 of Ref. 6). We don't
understand the reason for this large difference. In Fig. 3 we also show
the scaled data and the resulting linear fit at the smallest 
number density $n$ = $1 \times 10^{-7}$ we used in this study.
Our estimated critical temperature is $T_{c}/T^{0}_{c}$ = 1.0061(14).

The resulting dependence of the transition temperature (a)
$(\Delta T_{c}/T^{0}_{c})/an^{1/3}$ (b) $T_{c}/T_{c}^{0}$ on
$an^{1/3}$ is shown in Fig. 4. Whenever not shown, the error bars
in our simulation data are smaller than the symbol sizes. We also
show comparisons between our results (Filled circles), the PRL97
data (Open circles), and Eq. (\ref{arnold}) using Arnold's
results\cite{Arnold2} (Solid line). We see that the $(\Delta
T_{c}/T^{0}_{c})/an^{1/3}$ depends on the interaction strength
$an^{1/3}$ while the PRL97 data are considerably flatter and smaller
than our results (see Fig. 4(a)). Our calculated data approach to
Arnold's results. The linear behavior is seen only at small value
of $an^{1/3}$ (see Fig. 4(b)). Our calculated data are slightly
larger than Eq. (\ref{arnold}) using Arnold's
results\cite{Arnold2} (Solid line) in the linear region. The
dashed straight line is a fit to our data points for the four smallest
values of $n$. We found that the slope ($c_{1}$) is 1.32 $\pm$ 0.14.
This is in agreement with the recent MC values of three-dimensional 
$O(2)$ scalar $\phi^{4}$ field theory, $c_{1}$ = 1.29
$\pm$ 0.05 (Ref. 24) and $c_{1}$ = 1.32 $\pm$ 0.02 (Ref. 16), and
the result by variational perturbation theory (VPT), $c_{1}$ =
1.27 $\pm$ 0.11 (Ref. 29).

In summary, we have determined the interaction-induced shift of
the phase transition temperature for Bose-Einstein condensation of
homogeneous weakly interacting Bose gases in three dimensions
using PIMC and finite-size scaling. Our results show that
$\{(T_{c}-T_{c}^{0})/T_{c}^{0}\}/(an^{1/3})$ depends on the
interaction strength $an^{1/3}$ while the previous PIMC
results\cite{Gruter} are considerably flatter and smaller than our
results. We obtain $c_{1}$ = 1.32 $\pm$ 0.14, in agreement with
results from recent Monte Carlo methods of three-dimensional $O(2)$ 
scalar $\phi^{4}$ field theory\cite{Kashurnikov, Arnold2}
and variational perturbation theory\cite{Kastening2}.

\begin{acknowledgments}

We are greatly indebted to Professor P. Stancil for his critical reading of the manuscript and profound comments and 
Shan-Ho Tsai and H. K. Lee for helpful
discussions. This work was partially supported by NASA grant No.
NAG8-1771.

\end{acknowledgments}

\end{document}